\documentclass[twoside,agupp]{aguplus} % PREPRINT
\usepackage{graphicx}
\usepackage{psfrag}
\printfigures
\doublecaption{35pc}
\sectionnumbers% turn on section numbers
\lefthead{HUANG ET AL.}
\righthead{LOMA PRIETA LOG-PERIODICITY}

\received{December 23, 1999}
\revised{June 12, 2000}
\accepted{?}
\journalid{JGRB}{January 2001}
\articleid{n}{m}
\paperid{?}
\ccc{0000-0000/00/?\$05.00}
\cpright{AGU}{2000}

\authoraddr{Y. Huang and H. Saleur, Department of physics, University
  of Southern California, Los Angeles, California 90089-0484. (e-mail:
  yueqiang@earth.usc.edu; saleur@physics1.usc.edu)}

\authoraddr{D. Sornette, Institute of Geophysics and Planetary
  Physics, University of California, Los Angeles, California
  90095-1567. (e-mail: Didier.Sornette@unice.fr)}

\begin{document}

\title{Re-examination of log-periodicity observed in the seismic precursors of
  the 1989 Loma Prieta earthquake}

\author{Y. Huang\altaffilmark{1} and H. Saleur}
\affil{Department of Physics, University of Southern California, Los
  Angeles}

\author{D. Sornette\altaffilmark{2,3}} 
\affil{Institute of Geophysics and Planetary Physics, University of
  California, Los Angeles}

% address change
\altaffiltext{1}{Department of Earth Sciences, University of Southern
California, Los Angeles, California 90089-0740.}

% alternative addresses

\altaffiltext{2}{Department of Earth and Space Sciences, University of
  California, Los Angeles, California 90095-1567.}

\altaffiltext{3}{Laboratoire de Physique de la Matiere
  Condensee, CNRS UMR 6622 and Universit\'e de Nice-Sophia Antipolis,
  Nice, France.}

\begin{abstract}
Based on several empirical evidence, a series of papers has advocated the
concept that seismicity prior to a large earthquake can be understood in
terms of the statistical physics of a critical phase transition.  In this
model, the cumulative seismic Benioff strain release $\epsilon$ increases
as a power-law time-to-failure before the final event.  This power law
reflects a kind of scale invariance with respect to the distance to the
critical point: $\epsilon$ is the same up to a simple rescaling
$\lambda^z$ after the time-to-failure has been scaled by a factor
$\lambda$. A few years ago, on the basis of a fit of the cumulative
Benioff strain released prior to the 1989 Loma Prieta earthquake,
\citet{sornette199505} proposed that this scale invariance could be
partially broken into a discrete scale invariance,  defined such that the
scale invariance occurs only with respect to specific integer powers of a
fundamental scale ratio. The observable consequence of discrete scale
invariance takes the form of log-periodic oscillations decorating the
accelerating power law. They found that the quality of the fit and the
predicted time of the event are significantly improved by the
introduction of log-periodicity. Here, we present a battery of synthetic
tests performed to quantify the statistical significance of this claim.
We put special attention to the definition of synthetic tests that are as
much as possible identical to the real time series except for the
property to be tested, namely log-periodicity. Without this precaution,
we would conclude that the existence of log-periodicity in the Loma
Prieta cumulative Benioff strain is highly statistically significant. In
contrast, we find that log-periodic oscillations with frequency and
regularity similar to those of the Loma Prieta case are very likely to be
generated by the interplay of the low pass filtering step due to the
construction of cumulative functions together with the approximate power
law acceleration. Thus, the single Loma Prieta case alone cannot support
the initial claim and additional cases and further study are needed to
increase the signal-to-noise ratio if any. The present study will be a useful
methodological benchmark for future testing of additional events when the
methodology and data to construct reliable Benioff strain function become
available.
\end{abstract}

\begin{article}

\section{Introduction}
\label{lomaintro}

The idea that earthquakes are somewhat analogous to critical phenomena
of statistical mechanics has been gaining ground in the last few decades
\citep{chelidze1982, allegre1982, sornette1990, tumarkin1992,
  sornette199505, newman9511,bowman199810,jaume99:_evolv}.
One of the consequences of this new point of view is that events
occurring even several decades before a large main shock can be
considered as seismic precursors, and that a study of this precursory
seismicity might give a fairly good indication of when the impending major
earthquake will take place, and how big it is going to be.  Although
such considerations are still in infancy, and the usual caveats about
earthquake prediction must be kept in one's mind, a lot of
work has already been devoted to ``post-diction'', sometimes with
impressive success. One of the pioneering cases in that direction was
the 1989 Loma Prieta earthquake, where \citet{sornette199505} proposed
to see the empirical power law used by \citet{bufe199306} in the
perspective of criticality in the sense of 
statistical physics \citep{sorbook}. They found that the cumulative Benioff
strain starting about 50 years ago could be well-fitted with a power
law, giving rise to a post-diction for the main event of
($1990.3\pm4.1$), a reasonably satisfactory result.

Things got even more exciting after the paper of
\citet{sornette199505} where it was pointed out that the strong
oscillations around the power law could be fitted as well using a
complex exponent correction to scaling:
\begin{equation}
  \label{eq:epsilon-t}
  \epsilon(t)=A+B(t_f-t)^z\{1+C \cos[\omega\log(t_f-t)+\phi]\}\,.
\end{equation}
In this formula, the parameter $t_f$ is the time of the main
shock (a pure power law would correspond to $C=0$), and the best fit
gave rise to an estimate $t_f=1989.9\pm0.8$, considerably closer to 
the real date than the one in \citep{bufe199306}, and with much less
uncertainty. 

The existence of complex correction to scaling exponents could be
linked to an underlying discrete scale invariance, a very appealing
property from a theoretical point of view: the initial observation in
\citep{sornette199505} therefore spurred a lot of development
\citep{saleur199603a,saleur199608,johansen199610,huang199801,sornette1998pr,bowman199810,johansen99:_new_kobe,sammis9901,jaume99:_evolv}.

Although the quality of the fit in \cite{sornette199505} is, to the
naked eye, impressively good, the suspicion arose recently that the
oscillations in the Loma Prieta Benioff strain could be merely the
result of noise. The synthetic tests performed in
\citep{sornette199505} being somewhat incomplete, we have decided to
reanalyze this question much more carefully in the present work.

We have performed two types of synthetic tests to do this re-analysis.

In the first type, we consider random power laws (explained in Section
\ref{lomasyn1}), with parameters that
match those of the real data, and study whether noise can give rise to
log-periodic structures after integration.  The advantage of this
approach is that the two key ingredients of the analysis of the real
data (power law and integration) are captured in a simple way.  Its
drawback is that the synthetic data (a sequence of random numbers
drawn from a power law probability distribution, from which a
cumulative quantity is constructed) is not quite of the same nature
than the real data (a sequence of times and magnitudes, from which the
cumulative Benioff strain is constructed). In particular, the sampling
for the synthetic data is essentially periodic in log scale, and this
effect, combined with integration, is expected to give rise to
spurious log-periodic oscillations indeed.  Nevertheless, we find the
consideration of these synthetic tests quite useful, as it complement
the considerations in
\citep{huang99:_mechan_log_period_under_sampl_data}.

The second type of synthetic tests is devised especially to avoid this
issue of sampling: we generate data for both time and magnitude, in
such a way that the probability distributions of the synthetic
($t^s,m^s$) and the real ($t^r,m^r$) quantities are the same. There is
a problem in doing so: although the synthetic data and the real data
have the same distribution, there is no guarantee that the cumulative
Benioff strain constructed form the synthetic sequences is really a
power law, because a power law dependence involves higher-order
statistics (\emph{i.e.}, correlation and dependence) not captured by the
one-point distribution functions. To preserve the feature of the real
data that events are more frequent and with higher magnitudes when
closer to the main shock (or the last data point), we added a
reordering procedure which shuffles the synthetic sequences in such a
way that the event with the $j^{\mathrm{th}}$ magnitude is at the same
position as the real event with the $j^{\mathrm{th}}$ magnitude in the
real sequence.

We find that for both type of synthetic tests, it is, surprisingly,
highly possible to get spurious (that is, entirely due to noise)
log-periodic oscillations which are as good as those observed in the
real Loma Prieta data. This conclusion is made quantitative in a
variety of ways, in particular by studying the highest peak of the
spectrum of oscillations around the power law for the real data, and
building the probability distribution of such peaks for synthetic
data. We thus conclude that, at the present time, it is not possible
to distinguish the log-periodic oscillations observed in
\citep{sornette199505} from noise.

The present study is related to
\citep{huang99:_mechan_log_period_under_sampl_data}.  The common theme
is the investigation of the conditions under which log-periodicity can
be created spontaneously by noise. In
\citep{huang99:_mechan_log_period_under_sampl_data}, the goal is to
study in details the underlying mechanism, relying solely on the
manipulation of data: the generally found non-uniform sampling
together with a low pass filtering step, as occurs in constructing
cumulative functions, in maximum likelihood estimations and
de-trending, is enough to create apparent log-periodicity. A detailed
exploration of this mechanism has been offered in
\citep{huang99:_mechan_log_period_under_sampl_data} together with
extensive numerical simulations to demonstrate all its main
properties. It was shown that this ``synthetic'' scenario for
log-periodicity relies on two steps: 1) the fact that approximately
logarithmic sampling in time corresponds to uniform sampling in the
logarithm of time; 2) integration reddens the noise and, in a finite
sample, creates a maximum in the spectrum leading to a most probable
frequency in the logarithm of time.  In
\citep{huang99:_mechan_log_period_under_sampl_data}, this insight was
then use to to analyze the 27 best aftershock sequences studied by
\citep{kisslinger199107} and search for traces of genuine
log-periodic corrections to Omori's law, which states that the
earthquake rate decays approximately as the inverse of the time since
the last main shock. The observed log-periodicity was shown to almost
entirely result from the ``synthetic scenario'' due to the data
analysis. From a statistical point of view, resolving the issue of the
possible existence of log-periodicity in aftershocks will be very
difficult as Omori's law describes a point process with a uniform
sampling in the logarithm of the time. By construction, strong
log-periodic fluctuations are thus created by this logarithmic
sampling. In contrast, in the present paper, we apply the insight
obtained in \citep{huang99:_mechan_log_period_under_sampl_data}, to
study {\bf accelerated} power laws culminating in a finite-time
singularity at time $t_f$.

To be complete, we should also point out the following.
\citet{sornette199505} paper contains a forward prediction of an
earthquake in the Kommandorski Island region at time $t_f = 1996.3 \pm
1.1$ year. Forward predictions provide a much larger statistical
significance since the model parameters are estimated independently
and outside their domain of application (see below the discussion in
the section on the analysis procedure). Forward prediction has also
the quality of increasing the number of cases. The prediction of a
critical time is not enough, one must specify the magnitude of the
predicted earthquake. In \citep{sornette199505}, the magnitude was
not specified but can probably be taken following the specification of
\citet{bufe94:_alask} of a magnitude in the range 7.5-8.5 occurring in a
zone originally outlined by
\citet{nishenko91:_circum_pacif_seism_poten}.  The largest earthquake
in the Harvard catalog during the time period 1994-1998 has a moment
magnitude $M_W=6.6$ (1996/07/16, 56.16N, 164.98E). The same event has
the magnitude $M_S=6.4$ in the PDE catalog. If the prediction is
considered correct only if both its time and magnitude range is as
predicted, this prediction is a failure.  However, it is hard to draw
any firm conclusion based on this single case with respect to
usefulness of the critical earthquake concept and of log-periodicity
as the methodology has evolved significantly since the initial paper
of \citet{sornette199505} (see for instance
\citep{bowman199810,ouillon00}).

The plan of this paper is as follows. In Section
\ref{sec:analysis-procedure} we present some general observations on
synthetic tests and their interpretations. Details on our two types of
tests, together with their results, are presented in Sections
\ref{lomasyn1} and \ref{lomasyn2}. Our conclusions are collected in
Section \ref{sec:discussion-loma-syn}.

\section{Analysis Procedure}
\label{sec:analysis-procedure}

Usually, the major problem in establishing the statistical
significance of a forecasting procedure is its retrospective character
involving a limited number of data and a significant number of
explicit (the parameters of the fit) and implicit (the total time and
space windows used, etc.) degrees of freedom.  In such situations, the
calculation of statistical significance becomes very difficult and
uncertain as soon as the adjustable parameters are determined from the
data. The conclusions can then be artifacts of the processing
technique or of a selection bias.  The paper of \citet{sornette199505}
certainly suffers from this problem.  Since there are no general
methods or techniques that would allow us to overcome these
difficulties, we stick to a more modest approach, which turns out to
be sufficient to draw a clear and meaningful conclusion.

We develop two types of synthetic tests. For both types, the key part
is the comparison of the log-periodic oscillations in the real
seismicity to those in the synthetic sequences. Similar oscillations
should have similar frequencies and similar regularities, and this can
best be quantified by considering the spectrum of these oscillations.
We focus on the highest peak (characterized by angular frequency
$\omega$ and peak height $h$) in the spectrum, which quantifies the
most significant frequency component in the oscillations as a function
of $\log(t_f-t)$ (since we are looking for log-periodicity).  The Lomb
method \citep{press1992} is used instead of the usual Fourier
Transform, since the data points are not equidistantly spaced.

The ultimate purpose of synthetic tests is to get the significance
level (the probability of getting the same thing by accident) of the
real observation. One natural way of evaluating the significance level
would be, it seems, to count the number of synthetic peaks (defined as
spectral components with spectral power higher than neighboring
frequencies, in other words, local maximums in the spectrum) within the
intervals $\omega^s\in[\omega^r-\Delta,\omega^r+\Delta]$ and $h^s \ge
h^r$ (superscript $s,r$ denote synthetic and real data respectively),
and to normalize this by the total number of synthetic peaks. This
might however lead to incorrect conclusions. For instance, suppose
that, from 10000 synthetic peaks, only one peak with peak height
higher than 10 (the height of the real peak) and angular frequency
$\omega$ in the range of $[\pi,2\pi]$ was found: could we conclude
that the significance level is $0.01\%$ (or confidence level of
$99.99\%$)?  Probably not. To see why, suppose the distribution of
$\omega$'s were uniform in $[0,40]$, and the distribution in peak
height uniform in $[6.862, 10.004]$: then the probability of observing
one peak of height above 10 and $\omega \in [\pi,2\pi]$ would roughly
be 1/10000. But this small probability does not mean that the peak of
height above 10 and $\omega \in [\pi,2\pi]$ is highly significant!
Due to the uniform distribution in both $\omega$ and peak height, any
other pair of $(\omega,h)$ would have, in fact, been observed with the same
probability. To draw positive conclusions from such a simple counting
analysis, one would need to have in advance a theory indicating where
the peak should be found, and approximately at what height. It is not
clear that there is such a thing at present, and therefore we prefer
to use a more conservative approach.

In the language of statistics, this question is related to the 
difference between first-order and second-order statistics. 
In first-order statistics, we ask: ``what is the probability
to observe the peak we see just by chance?''  In second-order
statistics, we ask: ``what is the probability to observe some
(first-order significant) peak somewhere (whatever its position)?''
In our context, the determination of the confidence level 
within first-order statistics requires that we have an a priori
understanding of where the peak should be found. 

In the absence of any theoretical predictions, it seems natural, to
quantify the significance level of the log-periodic oscillations, to
rely somehow on the probability distribution of $\omega$ and peak
height in the synthetic samples, i.e. to rely on second-order 
statistics. We have not managed to come up with a
totally satisfactory, objective way to use this probability
distribution however. A useful quantity we came up with--but it
should not be trusted blindly--is the ratio $R$ of the probability of
observing a given peak to the probability of observing the most
probable peak: if the ratio is close to 1, this surely indicates that
the peak is not very significant.

To obtain this ratio, one needs the probability density function
$p(\omega,h)$ of the synthetic peaks: the latter can be constructed
using the Kernel Density method \citep{silverman1986,beardah1995}. We
then set $R=
\frac{p(\omega^r,h^r)\,d\omega\,dh}{p(\omega^{mp},h^{mp})\,d\omega\,dh}$
where $(\omega^r,h^r)$ characterize the real peak, and
$(\omega^{mp},h^{mp})$ is the most probable synthetic peak. The ratio
quantifies how frequently in synthetic sequences we can observe
log-periodicity similar to that observed in the real sequence. There
are some technical advantages in using this ratio. For instance, there
is in fact no arbitrariness in choosing the intervals $d\omega$ and
$dh$, since they cancel out between the numerator and the denominator:
it is then enough to just use the function value of $p(\omega,h)$ at
$(\omega^r,h^r)$ and $(\omega^{mp},h^{mp})$.  Also, the special
choices (e.g., the degree of smoothing and the type of kernels) made
in constructing the probability density function $p(\omega,h)$
hopefully also cancel out in the ratio, and have little influence on
the final result.

The generation of synthetic data and extraction of oscillations
depend on the type of synthetic tests and will be explained separately
in the following sections.

\section{Synthetic Test I}
\label{lomasyn1}

\subsection{Generation of Synthetic Data}

We start from a simple method which takes into account the most
crucial features of the real data. We then refine this method in
several ways in Section \ref{lomasyn2}.

The crucial features of the real data are power law and integration.
The latter--considering the cumulative Benioff strain--is necessary
for numerical reasons, as there are not enough data points to study
directly the rate at which energy is released (moreover, considering
the rate leads to other difficulties, like the influence of the
binning intervals etc). We therefore take a pure power law as the null
hypothesis, and generate data with a probability density fitting the
power law part of $d\epsilon/dt$: we mimic real data as closely as
possible, taking in particular the same number of points. We then
construct a synthetic cumulative Benioff strain by numerical
integration, and investigate whether noise in the sampling of the
power law can give rise to spurious log-periodic oscillations.

Again, the power law is 
\begin{equation}
  \label{eq:depsilont}
  \frac{d\epsilon(t)}{dt}\propto(t_f-t)^{m-1}
\end{equation}
following (1) in \citep{sornette199505}.  We assume the range for
$t$ is $[t_0,\ t_1]$ with $t_1<t_f$ to avoid the singularity at $t_f$.
After normalization, we have
\begin{equation}
  \label{eq:pw-loma}
  \frac{d\epsilon(t)}{dt}=\frac{m}{(t_f-t_0)^m-(t_f-t_1)^m}(t_f-t)^{(m-1)}.
\end{equation}
For the real seismic precursors, $t$ is the time of occurrence of an
earthquake, for synthetic events $t$ is a random variable with
probability function $p(t)=\frac{d\epsilon(t)}{dt}$. To make sure the
synthetic events and the real seismic precursors have the same power law
distribution, we chose the same  parameters $t_0,\ t_1,\ t_f,\ m,\ N$ 
 (from Table I and Figure 1 of \citep{sornette199505}), where
$N$ is the number of events. (Since the original data was not
  available at the time of this study, we retrieved data from the CNSS
  catalog using their space-time-magnitude window. However, due to
  some unknown reason, our data set was slightly different from their
  data set. We got only 27 events instead of 31.)  The random
variable $t$ with given $p(t)$ can be transformed from a random
variable $x$ uniformly distributed on $[x_0,\ x_1]$ by solving
\citep{press1992}
$$p(x)dx=p(t)dt.$$
The transformation is
\begin{eqnarray}
  \label{eq:tran}
  t&=&t_f-\left[(t_f-t_0)^{m} \right. \nonumber \\
   & &\left.-\frac{x-x_0}{x_1-x_0}\left((t_f-t_0)^{m}-(t_f-t_1)^{m}\right)\right]^{\frac{1}{m}}\,.
\end{eqnarray}
We then construct the cumulative distribution function of $t$ and use
this function to mimic the cumulative Benioff strain of the real
sequence. They have the same power law parameters and they are both,
indeed constructed from integration.

\subsection{Extraction of Oscillations}
\label{sec:extr-oscill-type-1}

The original analysis procedure used in \citep{sornette199505} was to
fit the cumulative Benioff strain to a power law with log-periodic
oscillations ((\ref{eq:epsilon-t}), the same as (8) in
\citep{sornette199505}).  The quality of the observed log-periodicity
was not quantified in \citep{sornette199505} other than by showing that
the quality of the fit measured by the residue as well as the
predicted critical time $t_f$ were both substantially improved
compared to those from the fit with the simple power law, but for our
study, it is crucial to do so.

To get the oscillations, we first fit a power law with log-periodic
oscillations (\ref{eq:epsilon-t}) to both the real and synthetic
data. The pure power law part (obtained by setting $C=0$ in
(\ref{eq:epsilon-t}) is then subtracted, and we obtain the
remaining oscillations.  These oscillations are in turn analyzed using
the procedure outlined in Section \ref{sec:analysis-procedure}.

\subsection{Results}
\label{sec:results-lp-syn1}

\subsubsection{The real sequence.}
\label{sec:real-sequence-1}

\begin{figure}
  \psfrag{t}[][]{$t$ (year)}
  \psfrag{cumulative Benioff strain}[][]{\ \ \ Normalized cumulative Benioff strain}
  \figbox*{\hsize}{}{\includegraphics{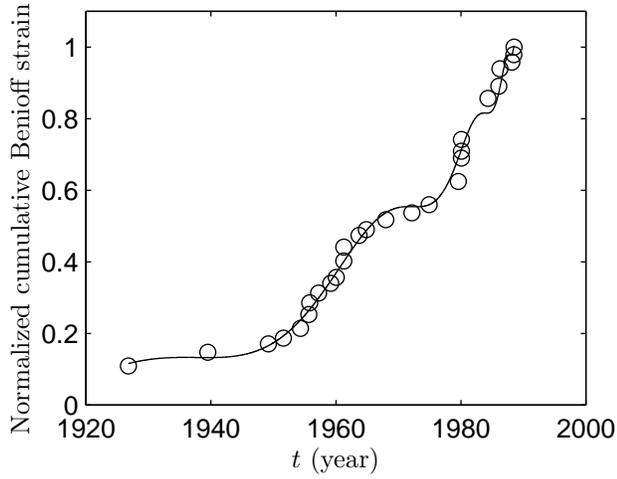}}
  \caption{The fit of a power law with log-periodic oscillations
    to the normalized cumulative Benioff strain of the seismic
    precursors of the 1989 Loma Prieta earthquake. }
  \label{fig:lpsyn1pwlga}
\end{figure}

\begin{figure}
  \psfrag{ - t}[][]{$t_f-t$ (year)}
  \psfrag{oscillations}[][]{oscillations}
  \figbox*{\hsize}{}{\includegraphics{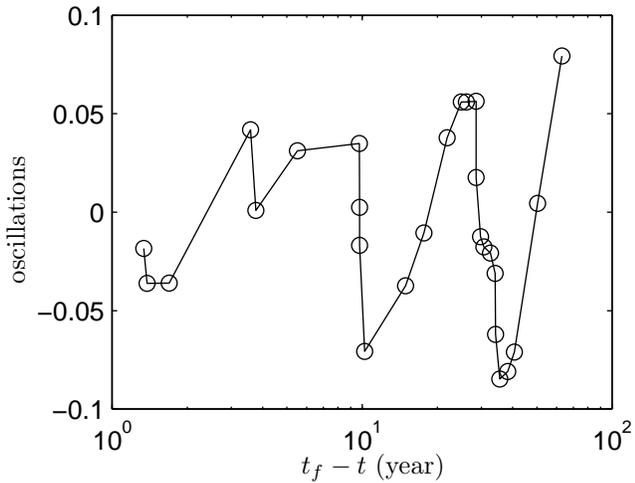}}
  \caption{The oscillations around the power law of 
    the normalized cumulative Benioff strain of the seismic
    precursors of the 1989 Loma Prieta earthquake. }
  \label{fig:lps1pwlgocsia}
\end{figure}

\begin{figure}
  \psfrag{w}[][]{$\omega$}
  \psfrag{Lomb Power}[][]{Lomb Power}
  \figbox*{\hsize}{}{\includegraphics{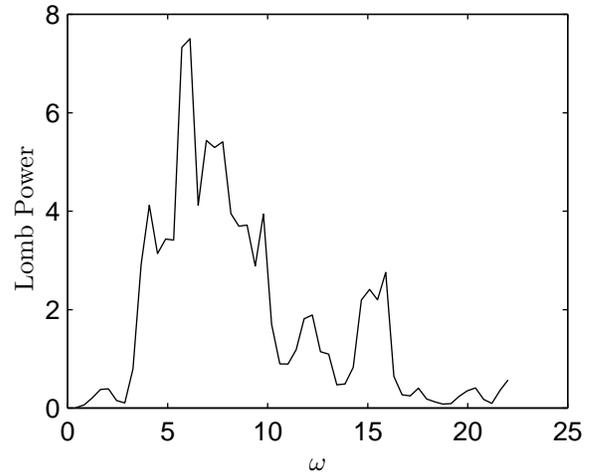}}
  \caption{The Lomb Periodogram of the oscillations shown in Figure
  \ref{fig:lps1pwlgocsia}. }
  \label{fig:lps1pwlgspeca}
\end{figure}

We first characterize the log-periodicity observed in the real
sequence. The fit of (\ref{eq:epsilon-t}) to the real data is
remarkable (Figure \ref{fig:lpsyn1pwlga}). The oscillations around the
power law part show approximately 2.5 cycles of regular oscillations
(Figure \ref{fig:lps1pwlgocsia}), the spectrum of which has a peak near
$\omega^r \sim 6.1$ with height $h^r\sim 7.5$ (Figure
\ref{fig:lps1pwlgspeca}), which is significantly different from
Gaussian noise (the chance of observing such a peak from
  Gaussian noise of the same number of data points is less than 2\%
  according to (13.8.7) of \citep{press1992}.). However, since we
are dealing with oscillations around a cumulative quantity, integrated
Gaussian noise would be a more appropriate null hypothesis. We will
study the chance of observing such a peak from integrated Gaussian
noise in Section \ref{sec:stat-from-synth-loma-syn-1}.
% figure 1
%
% figure 2
%
% figure 3

\subsubsection{Synthetic sequences.}
\label{sec:stat-from-synth-loma-syn-1}

\begin{figure}
  \psfrag{cumulative number of events}[][]{Cumulative number of
    events}
  \psfrag{t}[][]{$t$ (year)}
  \figbox*{\hsize}{}{\includegraphics{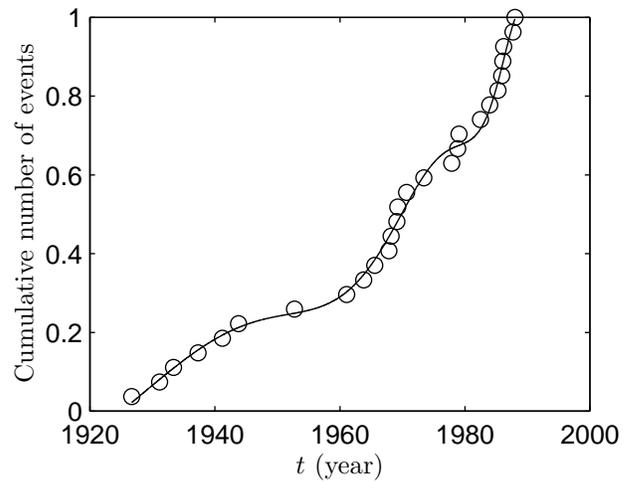}}
  \caption{Plot of a quantity similar to that in Figure
    \ref{fig:lpsyn1pwlga} from a synthetic sequence.}
  \label{fig:lpsyn1pwlgb}
\end{figure}

\begin{figure}
  \psfrag{oscillations}[][]{Oscillations}
  \psfrag{ - t}[][]{$t_f-t$ (year)}
  \figbox*{\hsize}{}{\includegraphics{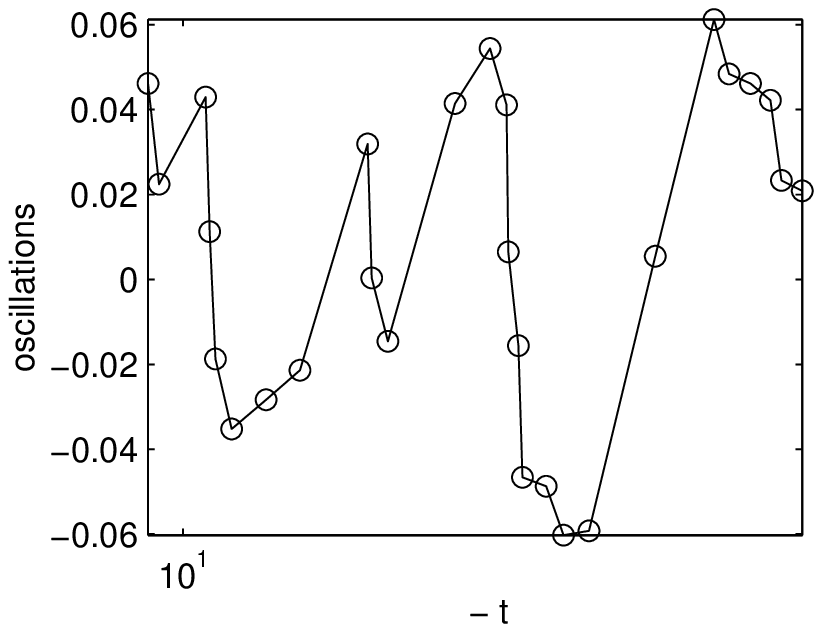}}
  \caption{Plot of a quantity similar to that in Figure
    \ref{fig:lps1pwlgocsia} from a synthetic sequence.}
  \label{fig:lps1pwlgocsib}
\end{figure}

\begin{figure}
  \psfrag{w}[][]{$\omega$}
  \psfrag{Lomb Power}[][]{Lomb Power}
  \figbox*{\hsize}{}{\includegraphics{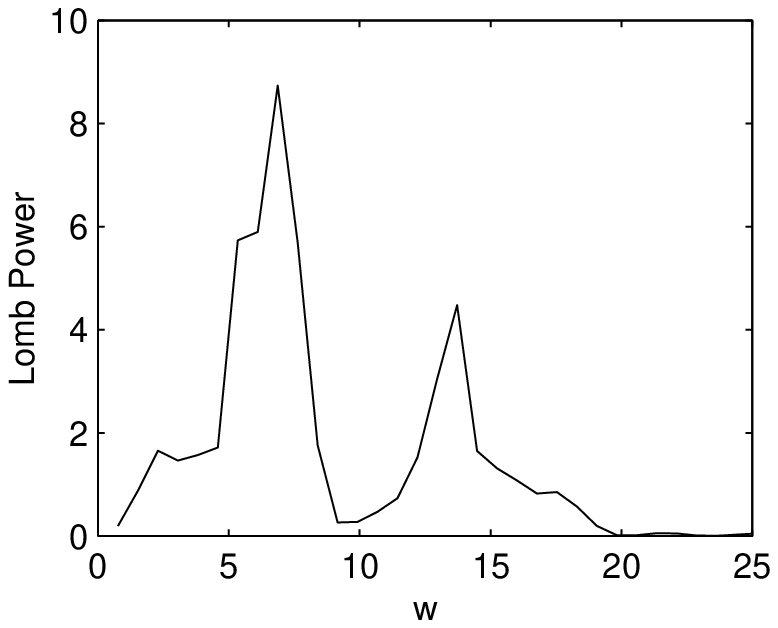}}
  \caption{Plot of a quantity similar to that in Figure
    \ref{fig:lps1pwlgspeca} from a synthetic sequence.}
  \label{fig:lps1pwlgspecb}
\end{figure}

300 synthetic\\ sequences were generated using the parameters of the
seismic precursors for the 1989 Loma Prieta earthquake. They 
 were analyzed in the same way as the real precursor
sequence. See Figures \ref{fig:lpsyn1pwlgb}, \ref{fig:lps1pwlgocsib},
and \ref{fig:lps1pwlgspecb} for typical results. We note that it is
possible to observe synthetic sequences which are remarkably similar
to the real sequence: similar amplitude of oscillations, similar
frequency, and similar regularity. In the following part, we quantify
how frequently such sequences can actually be observed.
%
% figure 4
%
% figure 5
%
% figure 6

\begin{figure}
  \psfrag{peak height}[][]{Peak height}
  \psfrag{w}[][]{$\omega$}
  \figbox*{\hsize}{}{\includegraphics{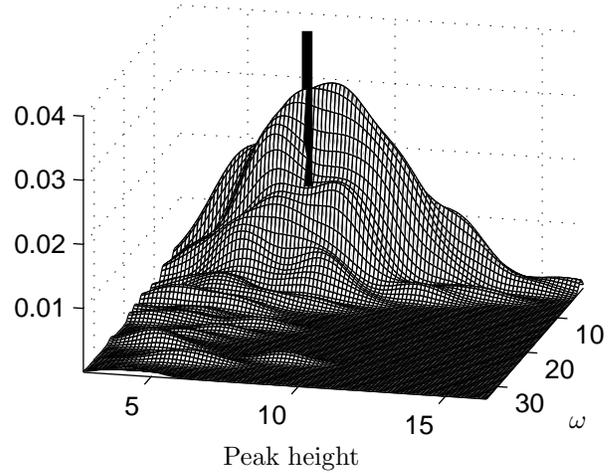}}
  \caption{The distribution function of frequencies 
    and peak heights from the synthetic sequences. The position of
    the real peak is marked by the vertical line. }
  \label{fig:lps1diswp3}
\end{figure}

The distribution function of the frequencies and peak heights of the
synthetic sequences were constructed using the Kernel Density method
\citep{silverman1986,beardah1995} (Figure \ref{fig:lps1diswp3}). The
real peak is not far from the most probable synthetic peak. From the
distribution function (Figure \ref{fig:lps1diswp3}), we find that the
probability density function at the most probable synthetic peak is
proportional to $0.031$, while it is proportional to $0.016$ at value
of $\omega$ and $h$ corresponding to the real peak. The ratio of these
two quantities is close to one half.  If we look at the separate
distribution of frequencies and peak heights (Figures
\ref{fig:lps1disw} and \ref{fig:lps1disp}), we see that the frequency
from the real sequence is slightly higher than the most probable
synthetic peak, and the peak height of the real sequence is almost the
most probable synthetic peak height. Note that the frequencies from
synthetic sequences have a rather narrow distribution, as expected
from \citep{huangphd,huang99:_mechan_log_period_under_sampl_data}.
The regularity of oscillations observed in the real sequence (Figure
\ref{fig:lpsyn1pwlga}, quantified by the peak height) is not surprising
due to the strong smoothing effect of integration \citep{huangphd,huang99:_mechan_log_period_under_sampl_data}.  When the same
analysis procedure was applied to both the real data and the synthetic
data, this kind of regularity is observed in both the real data and
the synthetic data.
%
% figure 7
%
% figure 8
\begin{figure}
  \psfrag{peak height}[][]{Peak height}
  \psfrag{w}[][]{$\omega$}
  \figbox*{\hsize}{}{\includegraphics{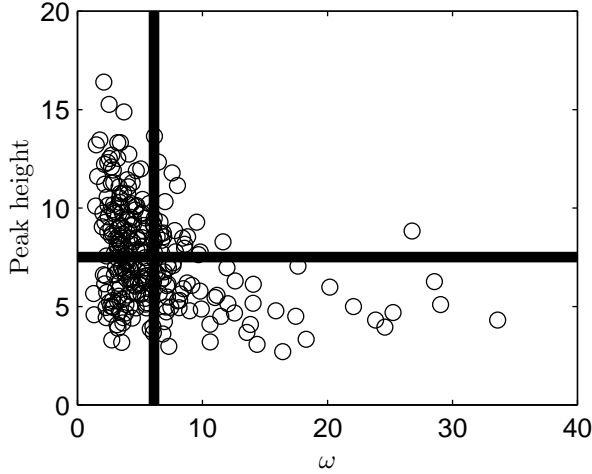}}
  \caption{2D map view of Figure \ref{fig:lps1diswp3}. Each circle
    represents one synthetic peak. The frequency of the real peak is
    marked by the vertical line, peak height by the horizontal line.}
  \label{fig:lps1diswp2}
\end{figure}
%
%figure 9
\begin{figure}
  \psfrag{w}[][]{$\omega$}
  \psfrag{probability density}[][]{Probability density}
  \figbox*{\hsize}{}{\includegraphics{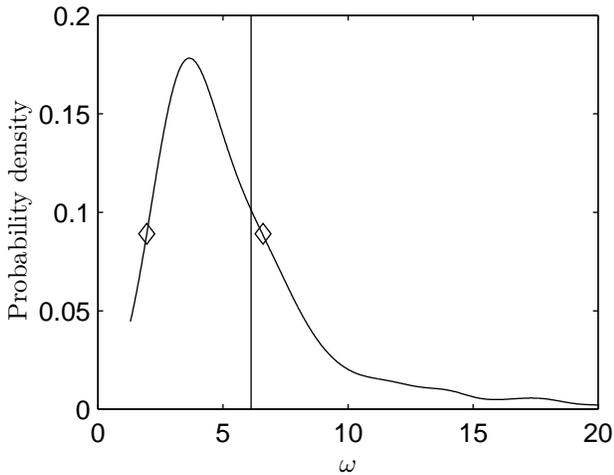}}
  \caption{The distribution of the synthetic frequencies. The
    vertical line marks the position of the real frequency. The two
    diamonds ($\diamond$) mark the FWHM (full-width-half-maximum) of the
    distribution (the same for all subsequent similar plots).}
  \label{fig:lps1disw}
\end{figure}
%
% figure 10
\begin{figure}
  \psfrag{peak height}[][]{Peak height}
  \psfrag{probability density}[][]{Probability density}
  \figbox*{\hsize}{}{\includegraphics{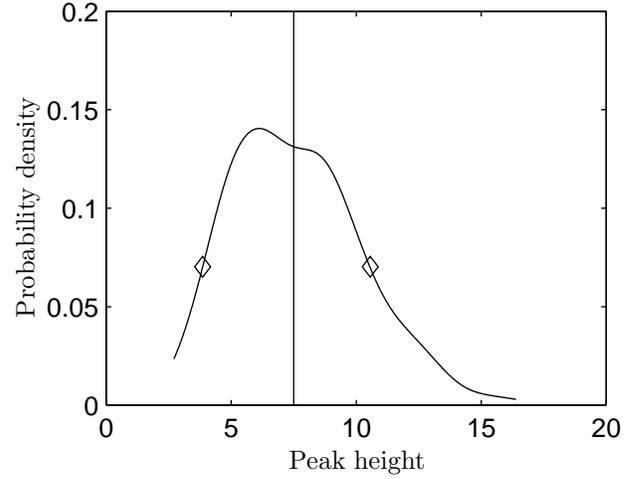}}
  \caption{The distribution of the synthetic peak
    heights. The vertical line marks the position of the real peak
    height. }
  \label{fig:lps1disp}
\end{figure}

\section{Synthetic Test II}
\label{lomasyn2}

\subsection{Generation of Synthetic Data}

For synthetic tests to be effective, the synthetic data should differ
from the real ones by only one characteristic--the characteristic to
be tested. Since in our problem we want to test log-periodicity in the
oscillations around a power law, the synthetic data should be the same
power law but with known noise (could be additional data
  errors or random fluctuations.).  Since the power law of the
cumulative Benioff strain is in fact implied by the magnitude
distribution and the temporal spacing between events and their
correlations, we have to generate synthetic magnitude $m^s$ and time
$t^s$ to get the same power law. $t^s$ and $m^s$ should be random
numbers having the same probability distribution as $t$ and $m$,
because in our observations we had control over neither $t$ nor $m$.
In this light, the analysis in Section \ref{lomasyn1} is thus a bit
oversimplified; we now refine it.

\begin{figure}
  \psfrag{t}[][]{$t$ (year)}
  \psfrag{cumulative distribution function}[][]{Cumulative distribution function}
  \figbox*{\hsize}{}{\includegraphics{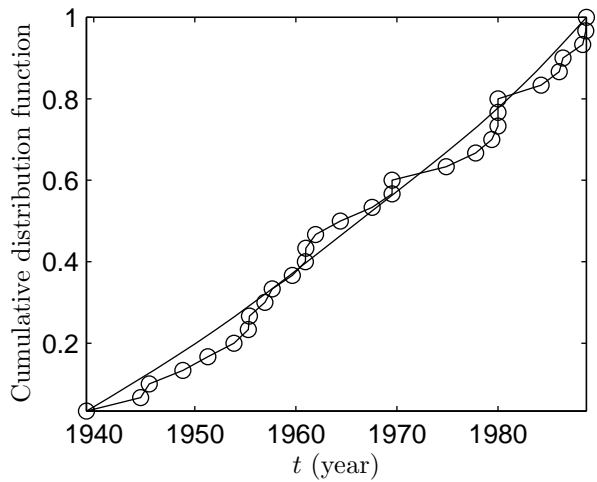}}
  \caption{The normalized cumulative number of events up to time t of
    the time sequence of the seismic precursors of the 1989 Loma
    Prieta earthquake (solid line connecting circles). The other solid
    line is the empirical line repeatedly smoothed (50 times) by
    3-point moving average.}
  \label{cumdis-t}
\end{figure}

One difficulty in generating synthetic $t^s$ and $m^s$ is that the
theoretical probability density function ($pdf$) of magnitudes and times
for our real data is unknown. This problem can be solved by using the
empirical $pdf$ constructed from the real data. However, we should not
use the empirical $pdf$ directly, otherwise all features of the real
data would be reproduced in the synthetic data.  For example, the
empirical distribution of the real time sequence (Figure
\ref{cumdis-t})
%
% figure 11
shows some regular oscillations (that could well be genuine
physically-based log-periodic oscillations) around its general trend.

If we used exactly this empirical distribution to generate the
synthetic time sequences, all synthetic sequences would show similar
regular oscillations: this is of course not appropriate, since what we
want to test is precisely whether these oscillations are the result of
noise. Therefore, we decided to use only the general trend of the
experimental data to generate our synthetic samples. This general
trend is obtained by smoothing the empirical distribution. Three-point
moving average was applied repeatedly 50 times (10 for smoothing the
cumulative distribution function of magnitudes).  The number of times
is not crucial as long as the oscillations were wiped out. The
criteria for our choice are closeness to the empirical curve and lack
of oscillations. Similar considerations were applied to the
generation of synthetic magnitude sequence. The assumption for the
general trend was not crucial. A reasonable one, close to the
empirical cumulative distribution curve but without the fluctuations,
would suffice.

\begin{figure}
  \psfrag{t}[][]{$t$ (year)}
  \psfrag{magnitude}[][]{Magnitude}
  \figbox*{\hsize}{}{\includegraphics{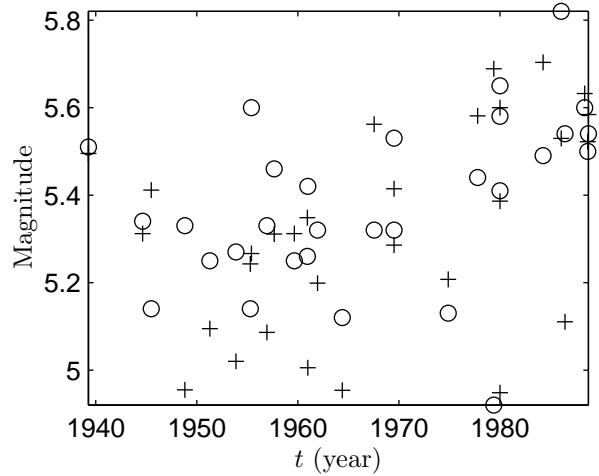}}
  \caption{Synthetic magnitude sequence (\texttt{+}). The real
    sequence is plotted with \texttt{o}.}
  \label{fig:lps3synm}
\end{figure}

\begin{figure}
  \psfrag{t}[][]{$t$ (year)}
  \psfrag{cumulative number}[][]{Cumulative number}
  \figbox*{\hsize}{}{\includegraphics{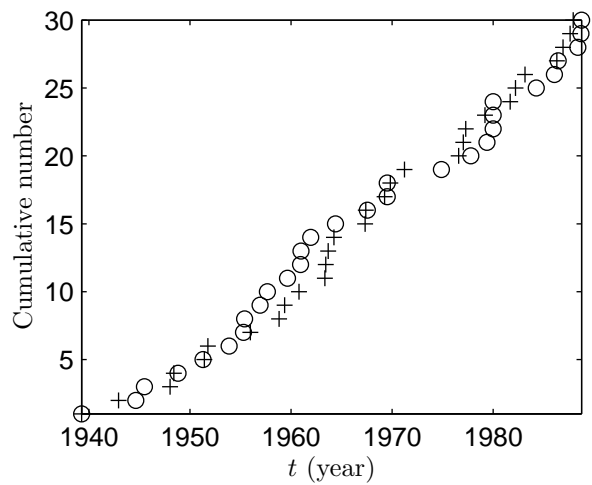}}
  \caption{Synthetic time sequence (\texttt{+}). The real sequence is
    plotted with \texttt{o}.}
  \label{fig:lps3synt}
\end{figure}

The next difficulty is that, for a sequence of times and magnitudes
generated using this method, there is no guarantee that the cumulative
Benioff strain will follow a power law. The time sequence more or less
follows a power law (we have verified that the cumulative
  number of events of the seismic precursors of the 1989 Loma Prieta
  earthquake is similar to the cumulative Benioff strain in power law
  and log-periodic oscillations), but, when combined with the
magnitude sequence to construct the whole Benioff strain curve, there
is no obvious reason why we should always get a power law.  It is
natural to expect that the power law of the real sequence comes mainly
from the fact that events occur more frequently with increasing
magnitude (trend only) when closer to the main shock. To preserve this
feature in the synthetic sequences, we decided to \bf{reorder} the
events in the synthetic sequence such that the event of the $k^{th}$
magnitude would occur at the same position in both the real and the
synthetic cases (for example, both the real sequence and the synthetic
sequences have the event of the second biggest magnitude being the
$k^{\mathrm{th}}$ one in the sequence).  This reordering scheme is
applied only to the magnitude sequence (time sequence is an ordered
sequence by definition).  One example is shown in Figures
\ref{fig:lps3synm} and \ref{fig:lps3synt}.
%
%figure 12
%
% figure 13

We performed synthetic tests both with and without the reordering
scheme. In fact, the results turned out to be almost identical. 

\subsection{Extraction of Oscillations}
\label{sec:extr-oscill-type-2}

The method of extracting oscillations from the cumulative Benioff
strain is slightly different from that of Section
\ref{sec:extr-oscill-type-1}, however the difference turns out to be
insignificant.

Two ways are possible to obtain the oscillations: the first involves
extracting the best-fit power law from the real data. The drawback of
this approach is that power law fits are often not as stable as fits
including the log-periodic corrections. Sometimes (around 8\% of all
cases) the fit even converges to a $t_f$ smaller than the time of the
last data point, thus $t_f-t$ is negative and $(t_f-t)^{(m-1)}$ is
complex since $m<1$, which is of course unphysical. The advantage of
this approach is that log-periodicity is not assumed in the first
place. The second approach involves extracting the power law obtained
from a best-fit power law with log-periodic oscillations
\citep{johansen9901a,johansen9901b,johansen9903}. The advantage here
is that the fit always converges well, but now log-periodicity is
somewhat assumed from the very beginning. The more positive view point
advocated in \citep{johansen9901a,johansen9901b,johansen9903} to
justify this procedure is that fitting with log-periodicity allows one
to take the most probably noise into account
\citep{huang99:_mechan_log_period_under_sampl_data} and thus to obtain
a good pure power law representation by putting the coefficient $C=0$.
In practice, the results of either approach were very similar: in the
following, we report only the results using the second one.

It is of course crucial to use exactly the same analysis procedure for
both the real data and the synthetic data, otherwise features
generated by the analysis procedure for the real data may not be
detected by the synthetic tests.

The cumulative Benioff strain was first constructed from the magnitude
sequence $m_i$:
\begin{equation}\label{bs}
  \epsilon_i=\sum\limits_{j=1}^i 10^{0.75m_j}
\end{equation}
and then normalized such that $\epsilon_{\mathrm{max}}=1$ (the unit was
changed without influencing the conclusion).  As in
\citep{sornette199505}, the cumulative Benioff strain was then fitted
to a power law with log-periodic oscillations.

The de-trended data were obtained by
\citep{johansen9901a,johansen9901b,johansen9903}
\begin{equation}
  \label{detren}
  detrn = \frac{\epsilon(t) - A}{B(t_f-t)^z}
\end{equation}
which should be either noise or pure log-periodic cosine according to
(\ref{eq:epsilon-t}), and then analyzed by the procedure explained
in Section \ref{sec:analysis-procedure}.

%There are 31 events in the original data used in
%\cite{sornette199505}. However the original data were not available to
%us in the format of time and magnitude. The data (kindly provided by
%A. Johansen and C. Sammis) were in the format of time and cumulative
%energy.  Thus to get the sequence of magnitude, the first data point
%had to be discarded (the cumulative energy was not calculated from the
%first event, but from many events before the first event)). The
%justification for discarding the first event was that the
%log-periodicity should not be significantly changed by excluding only
%one event if it was physically genuine. Also the inclusion or
%exclusion of several events due to uncertainties in magnitude was not
%very unusual. The magnitudes were inverted from energy starting from
%the second event using $E=10^{4.8+1.5m}$ which was supposed to be used
%in \cite{bufe199306} (private communication with David Bowman). In the
%original data the last event had a time smaller than the second last
%event, we exchanged them in this study. The magnitudes inverted this
%way seemed to be in reasonable agreement with the data set we
%retrieved from the CNSS catalog. Synthetic sequences of time and
%magnitude were generated according to the real sequence of time and
%magnitude.

\subsection{Results}\label{sec:results}

\subsubsection{The real sequence.}\label{sec:real-sequence}

The fitting of the real data
%
% figure 14
\begin{figure}
  \psfrag{time}[][]{$t$ (year)}
  \psfrag{Normalized Benioff strain}[][]{Normalized Benioff strain}
  \figbox*{\hsize}{}{\includegraphics{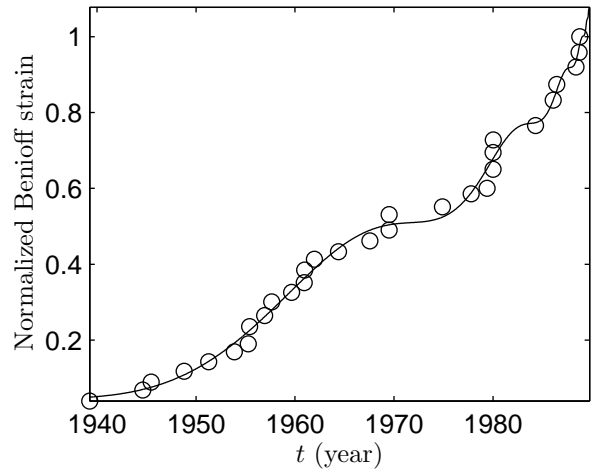}}
  \caption{The fitting of the normalized cumulative Benioff
    strain of the 30 seismic precursors of the 1989 Loma Prieta earthquake
    to (\ref{eq:epsilon-t}). }
  \label{lps_pwlg}
\end{figure}
showed good agreement between the real data and the theoretical curve
(Figure \ref{lps_pwlg}).
% 
% figure 15
\begin{figure}
  \psfrag{c}[][]{$\log(t_f-t)$}
  \psfrag{de-trended Benioff strain}[][]{De-trended Benioff strain}
  \figbox*{\hsize}{}{\includegraphics{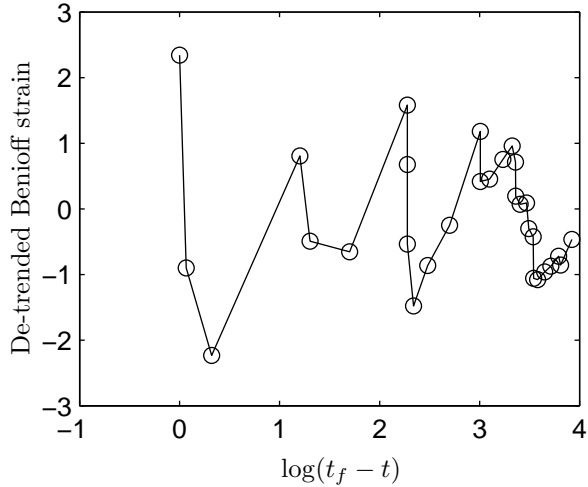}}
  \caption{The de-trended data of the normalized Benioff
    strain of the 30 seismic precursors of the 1989 Loma Prieta
    earthquake.}
  \label{lps_osci}
\end{figure}
%
% figure 16
\begin{figure}
  \psfrag{w}[][]{$\omega$}
  \psfrag{Normalized Lomb density}[][]{Normalized Lomb density}
  \figbox*{\hsize}{}{\includegraphics{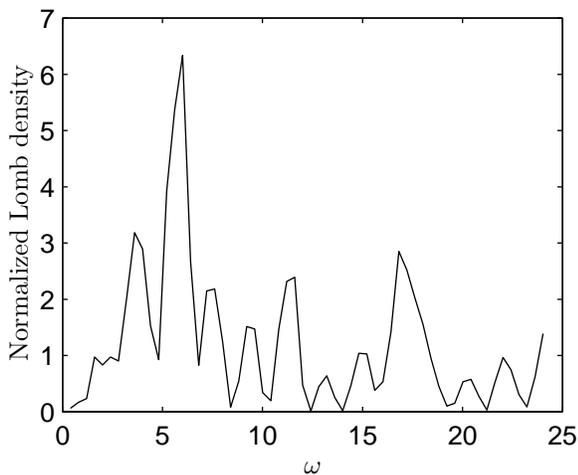}}
  \caption{Spectrum of the de-trended data in Figure \ref{lps_osci}.}
  \label{lps_spectrum}
\end{figure}
There was a peak at $\omega=6.1$ in the spectrum of the de-trended
data (Figure \ref{lps_spectrum}), close to the best-fit parameter
$\omega=5.7$. 

\subsubsection{Synthetic sequences.}
\label{sec:with-reordering}

\begin{figure}
  \psfrag{t}[][]{$t$ (year)}
  \psfrag{Normalized Benioff strain}[][]{Normalized Benioff strain}
  \figbox*{\hsize}{}{\includegraphics{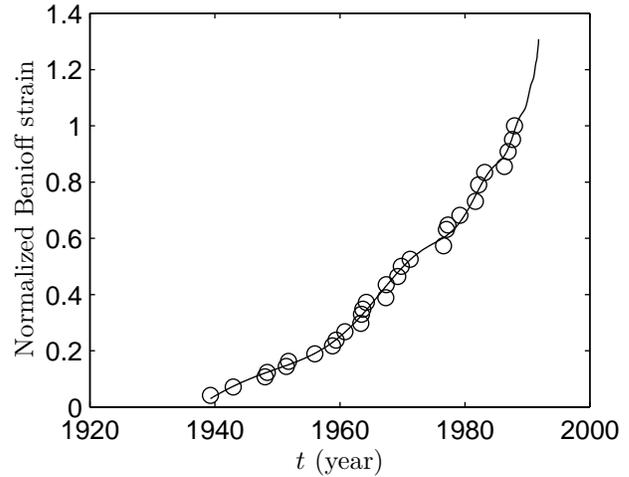}}
  \caption{The fitting of the normalized cumulative Benioff
    strain of one synthetic sequence
    to (\ref{eq:epsilon-t}).}
  \label{lps_pwlgsyn}
\end{figure}

1000 synthetic sequences (one example in Figures \ref{fig:lps3synm}
and \ref{fig:lps3synt}) were analyzed using the same procedure as that
for the real sequence.  The cumulative Benioff strain of the synthetic
sequence showed obvious similarity to the real data (Figure
\ref{lps_pwlgsyn}).
%
% figure 17

We first checked whether the synthetic data gave rise to power laws.
The method to do this was to fit these data to a power law shape, and
measure the summed square of error (SSE) from the fitting: in about
50\% of the cases, the synthetic sequences had an SSE smaller than
that of the real sequence, indicating that they were roughly as good
power laws as the real data.

We note here that the ratio of the probability of observing a
synthetic sequence with SSE similar to that from the real sequence
divided by the probability of observing a synthetic sequence with the
most probable SSE is around 79\%. If we used SSE to measure the
goodness of the power law model for our synthetic data, the real
sequence would be very close to the most probable synthetic sequence.

We then compared the fit of a power law with log-periodic oscillations
to the real and the synthetic sequences.

The SSE of the synthetic sequences are in the range [0.005, 0.05],
centered around the SSE from the real sequence ($\sim 0.013$). There
are around 1/4 of the synthetic sequences that have a SSE smaller than
that from the real sequence. The ratio of the probability of observing
a synthetic sequence with SSE similar to that from the real sequence
and the probability of observing a synthetic sequence with the most
probable SSE is around 96\%. If we use SSE to measure the regularity
of log-periodic oscillations, the real sequence is thus very close to
the most probable synthetic sequence.

\begin{figure}
  \psfrag{tc}[][]{$t_f$ (year)}
  \psfrag{probability density}[][]{Probability density}
  \figbox*{\hsize}{}{\includegraphics{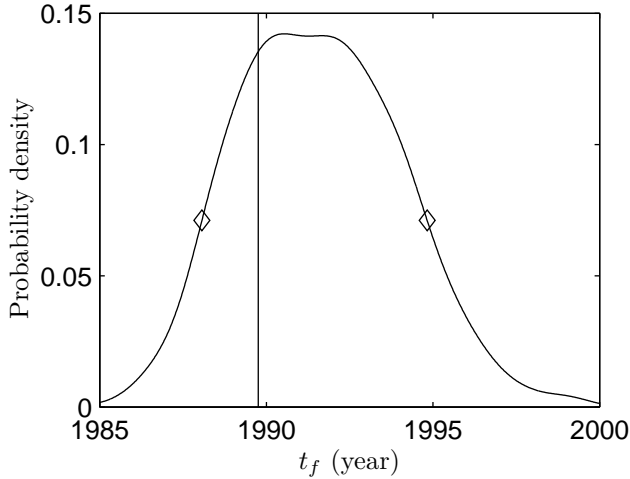}}
  \caption{The distribution of main shock times ($t_f$ in
    (\ref{eq:epsilon-t})) from the synthetic sequences.  The
    vertical line marks the value from the real sequence.}
  \label{fig:lpssynpwlgdistc}
\end{figure}

The $t_f$ from the synthetic sequences is distributed in [1988,1995]
(Figure \ref{fig:lpssynpwlgdistc}). The ratio of the probability of
observing a synthetic sequence with $t_f$ similar to that from the
real sequence divided by the probability of observing a synthetic
sequence with the most probable $t_f$ is around 95\%. Thus the
apparent accurate prediction of the actual main-shock time for the
1989 Loma Prieta earthquake \citep{sornette199505} might be due to
chance. The width of the distribution (FWHM, full width at half
maximum) is 6.8 years, narrower than that from the fit of a pure power
law (7.8 years), suggesting that log-periodicity may improve 
power law fits by accounting for the most probable noise
\citep{huang99:_mechan_log_period_under_sampl_data}.
%
% figure 18

\begin{figure}
  \psfrag{w}[][]{$\omega$}
  \psfrag{probability density}[][]{Probability density}
  \figbox*{\hsize}{}{\includegraphics{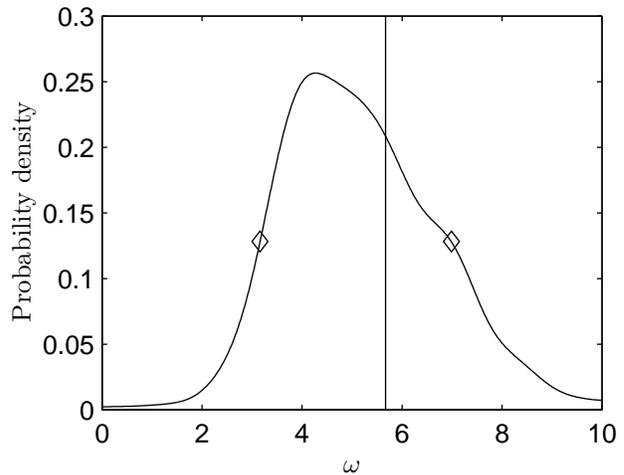}}
  \caption{The distribution of frequencies of log-periodic
    oscillations from the fitting of the synthetic sequences. The
    vertical line marks the value from the real sequence.} 
  \label{fig:lpssynpwlgdisw}
\end{figure}

There is a well-defined peak in the distribution of the frequencies of
the log-periodic oscillations from the fitting of the synthetic
sequences (Figure \ref{fig:lpssynpwlgdisw}). The ratio of the
probability of observing a synthetic sequence with frequency similar
to that from the real sequence divided by the probability of observing
a synthetic sequence with the most probable frequency is around 81\%.
%
% figure 19

We summarize the above results in Table \ref{tab:para-loma-syn}.
\begin{table*}
  \tablewidth{35pc}
  \caption{Parameters From the Fit of the Synthetic Data With PW
    (Pure Power Law) and PWLG (Power Law With Log-Periodic
    Oscillations)\label{tab:para-loma-syn}}
  \begin{center}
    \begin{tabular}{cccccccccc}\\[-3ex]\hline\\[-1ex]
        %-----------------------------------------------------------------------------------------------
        &       & pr\tablenotemark{\it a}\tablenotetext{\it a}{Value of the probability density
                function at a synthetic peak similar to the real peak.} 
                & pmp\tablenotemark{\it b}\tablenotetext{\it b}{Value of the probability density
                function at the most probable synthetic peak.}  
                & ratio   & mp\tablenotemark{\it c}\tablenotetext{\it c}{The most probable value from synthetic data.}    
                          & r\tablenotemark{\it d}\tablenotetext{\it d}{The value from real data.}      
                & left\tablenotemark{\it e}\tablenotetext{\it e}{Value at the left point of the FWHM of the distribution of synthetic values.}  
                & right\tablenotemark{\it f}\tablenotetext{\it f}{Value at the right point of the FWHM of the distribution of synthetic values.} 
                & FWHM\tablenotemark{\it g}\tablenotetext{\it g}{Full width at Half Maximum of a peak.}\\[1ex]\hline\\[-1ex]
        %-----------------------------------------------------------------------------------------------
        & m     & 0.72  & 1.99  & 0.36  & 0.34  & 0.69  & 0.17  & 0.57  & 0.40 \\ 
PW      & tc    & 0.073  & 0.13  & 0.56  & 1994.3 & 1988.7 & 1988.4 & 1996.2 & 7.8    \\ 
        & SSE   & 20.0 & 24.9 &  0.79 & 0.028  & 0.037  & 0.016  &
        0.052  & 0.036 \\[1ex]\hline\\[-1ex]
        %-----------------------------------------------------------------------------------------------
        & m     & 1.94  & 2.23  & 0.87  & 0.45  & 0.52  & 0.25  & 0.63  & 0.38 \\ 
        & tc    & 0.13  & 0.14  & 0.95  & 1990.5 & 1989.8 & 1988.1 & 1994.8 & 6.8    \\
PWLG    & SSE   & 50.0 & 51.8 & 0.96  & 0.015  & 0.017  & 0.0092  & 0.026  & 0.014 \\
        & w     & 0.21  & 0.26  & 0.81  & 4.28  & 5.67  & 3.16  & 6.99  & 3.83 \\
        & C     & 2.84  & 6.71  & 0.42  & -0.039 & -0.083 & -0.078 & 0.075  & 0.15 \\[1ex] \hline\\[-3ex]
        %-----------------------------------------------------------------------------------------------
    \end{tabular}
    \end{center}
\end{table*}

\begin{figure}
  \psfrag{p}[][]{Peak height}
  \psfrag{probability density}[][]{Probability density}
  \figbox*{\hsize}{}{\includegraphics{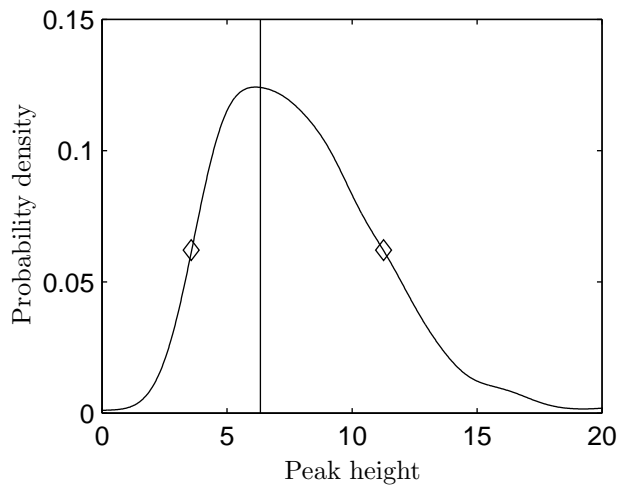}}
  \caption{The distribution of peak heights from the
    de-trended data. The vertical line marks the value from the
    real sequence.}
  \label{fig:lpssynwdisp}
\end{figure}

We now turn to the statistics from the characterization of log-periodicity
by spectral analysis of the de-trended data.  If we only look at the
distribution of peak heights, the ratio of the probability observing a
synthetic peak similar to the real peak divided by the probability of
observing the most probable synthetic peak is around 99.9\% (Figure
\ref{fig:lpssynwdisp}).
%
% figure 20

\begin{figure}
  \psfrag{w}[][]{$\omega$}
  \psfrag{probability density}[][]{Probability density}
  \figbox*{\hsize}{}{\includegraphics{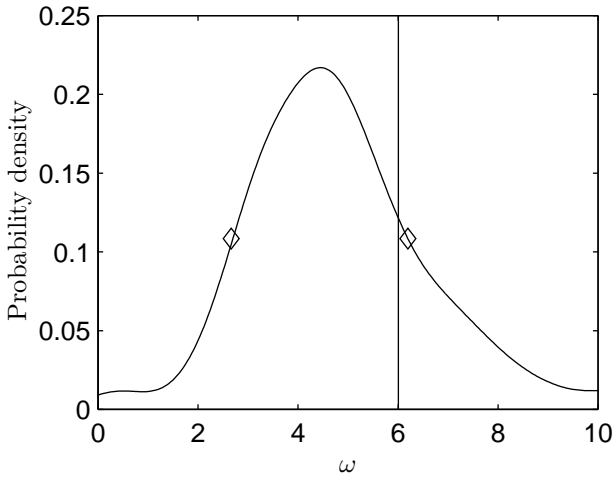}}
  \caption{The distribution of frequencies from
    the de-trended data.  The vertical line marks the value from the
    real sequence.}
  \label{fig:lpssynwdisw}
\end{figure}

If we only look at the frequencies, the ratio of the probability of
observing a synthetic frequency similar to that of the real peak
divided by the probability of observing the most probable synthetic
frequency is around 56\% (Figure \ref{fig:lpssynwdisw}).
%
% figure 21

\begin{figure}[htbp]
  \psfrag{w}[][]{$\omega$}
  \psfrag{peak height}[][]{Peak height}
  \figbox*{\hsize}{}{\includegraphics{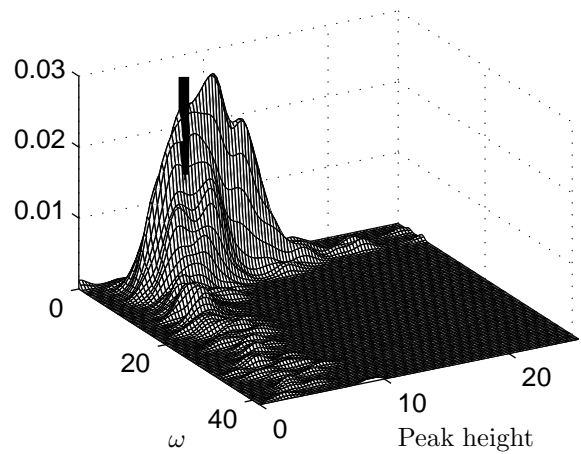}}
  \caption{The distribution of peak heights and frequencies
    from the de-trended data.}
  \label{fig:lpssynwdispw3}
\end{figure}

\begin{figure}[htbp]
  \psfrag{w}[][]{$\omega$}
  \psfrag{peak height}[][]{Peak height}
  \figbox*{\hsize}{}{\includegraphics{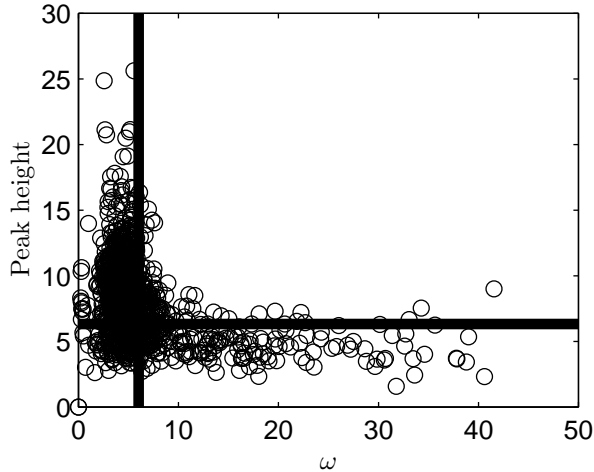}}
  \caption{The 2D map view of Figure \ref{fig:lpssynwdispw3}.  The vertical
    and horizontal lines mark the values from the real sequence.}
  \label{fig:lpssynwdispw2}
\end{figure}

When we look at the joint distribution of peak heights and
frequencies, the ratio of the probability for  observing a peak
similar to the real peak divided by the probability of 
observing the most probable synthetic peak is about 56\%
(Figures \ref{fig:lpssynwdispw3} and \ref{fig:lpssynwdispw2}).
%
% figure 22
%
% figure 23

Not using the reordering scheme did not significantly change the above
results, except that, for some synthetic sequences, the power law was
not good.

Also, recall that the foregoing results involved fitting data to a
power law with log-periodic oscillations, then de-trending.  Fitting
data to a pure power law instead produced very similar results.

We summarize all the results in the previous sections in Table
\ref{tab:loma-prieta-syn}.
\begin{table}[htbp]
  \caption{Synthetic Tests of the Log-Periodicity Observed in the
    Benioff Strain of the Seismic Precursors of the 1989 Loma Prieta
    Earthquake}
  \label{tab:loma-prieta-syn}\vskip 0.5ex
  \begin{center}
    \begin{tabular}{ccccccc}\\[-5ex]\hline\\[-1ex]
&\multicolumn{3}{c}{EOPW\tablenotemark{\it a}\tablenotetext{\it a}{Extracted oscillations using the best-fit pure power law.}}
& \multicolumn{3}{c}{De-trended data}\\[1ex]\hline\\[-1ex]
& pr\tablenotemark{\it b}\tablenotetext{\it b}{Value of the probability density function at a synthetic peak similar to the real peak.}
& pmp\tablenotemark{\it c}\tablenotetext{\it c}{Value of the probability density function at the most probable synthetic peak.} 
& ratio & pr & pmp & ratio \\[1ex]\hline\\[-1ex] 
        $\omega$ & 0.11& 0.11& 0.99 &0.12 &0.22 &0.56 \\
         $h$& 0.19 &0.20  &0.93  & 0.12&0.12 &1.00 \\
        ($\omega$, $h$) &0.027 &0.031 &0.89 &0.016 &0.029 &0.56  \\[1ex]\hline\\[-1ex]
        ($\omega$, $h$)\tablenotemark{\it d}\tablenotetext{\it d}{No reordering.} & 0.020 & 0.025 & 0.81 & 0.16& 0.23& 0.73 \\[1ex]\hline\\[-3ex]
      \end{tabular}
    \end{center}
\end{table}

\section{Discussion}
\label{sec:discussion-loma-syn}

These synthetic tests were  performed to determine whether it is
possible to observe the reported \\log-periodicity \citep{sornette199505}
from integrated noise in power laws, and if possible, how big the
probability is. We found that, if we use the highest peak of the
spectrum of the oscillations around the power law of the cumulative
Benioff strain to quantify the log-periodicity in the oscillations,
peaks similar to the peak observed from the real sequence (the real
peak) were indeed frequently observed from the synthetic sequences.
The odds of observing a synthetic peak similar to the real peak is
more than 50\% of the odds of observing the most probable synthetic
peak.

It is reasonable to use the highest peak in the spectrum of a signal
to quantify the most significant frequency component in the signal.
The position of the peak is the frequency ($\omega$), and the peak
height ($h$) quantifies the regularity of that frequency component. If
two signals have similar peaks in their spectrums, they must have
oscillations of similar frequency and regularity.

Peaks similar to the real peak were observed frequently from the
synthetic sequences. To quantify this frequency of observation, we
constructed the probability density function $p(\omega,h)$ of the
synthetic peaks in the space of $(\omega,h)$. From $p(\omega,h)$, we
were able to obtain the probability of observing a synthetic peak
similar to the real peak $(\omega^r,h^r)$, that is
$p(\omega^r,h^r)\,d\omega\,dh$. This probability might be a small
number, which is meaningful only when compared with the probability of
observing the most probable synthetic peak, which is
$p(\omega^{mp},h^{mp})\,d\omega\,dh$. The ratio of the two
probabilities quantifies well how frequently we observe the feature
from synthetic data.

The mechanism at the origin of log-periodicity in the synthetic data
sets has been discussed in
\citep{huangphd,huang99:_mechan_log_period_under_sampl_data}. Briefly,
log-periodicity results from the fact that taking the cumulative of a
power law involves a low pass filtering step (reddening of the noise)
which, in a finite sample, creates a maximum in the spectrum leading
to a most probable log-frequency corresponding approximately to $1.5$
cycles over the full sampled interval.

We looked into two quantities for log-periodicity in the oscillations
around the power law of the cumulative Benioff strain. The extracted
oscillations are the difference between the data and the best-fit
power law. The de-trended data were obtained using (\ref{detren}).
For both of them, the ratio of the two probabilities is bigger than
50\%, which means that it is not only possible to observe that kind of
log-periodicity in synthetic data, but also highly probable.

Our synthetic events and the real events have the same distribution in
time and magnitude, and they were analyzed in exactly the same way.
Since discrete scale invariance is not present in the synthetic data,
the log-periodicity observed in the real sequence  cannot
be used as evidence for  discrete scale invariance.

In fact, even the power law--used as evidence of ordinary scale
invariance--could also be explained by other mechanisms. Indeed, the
cumulative Benioff strains of the synthetic sequences in Section
\ref{lomasyn2} do follow power laws similar to that of the real
sequence. For the real sequence, the cumulative distribution function
of event times is not significantly different from a straight line.
The cumulative distribution of moments is not a power law either ($S$
shape instead of the usual power law shape) (this is an
  \bf{ad hoc} statement. The reason is that for this sequence of
  magnitudes, the number of data points is small (only 30) and the
  magnitude cut-off is very high (5.0). So even if in general the
  moment distribution is a power law, large statistical fluctuations
  may make the moment distribution of this sequence non power law. We
  use an empirical distribution function instead of the usual power
  law assumption to avoid dependence on that assumption. In fact our
  method will still be valid no matter what the underlying
  distribution is.).  But for the real sequence, magnitudes tends to
be bigger when closer to the main shock, especially for the last
several events \citep{Jones}. Since small difference in magnitudes
will be translated into quite big difference in the cumulative Benioff
strain, the last several events would make the would-be linear trend
bend upward which happens to be well described by a power law of small
exponent. Since this increasing tendency of magnitudes was preserved
in the generation of synthetic magnitudes, power laws are also good
for describing the synthetic data. In fact, without the re-ordering
scheme to preserve that feature of magnitudes, we were still able to
obtain similar results.  As long as the magnitudes are not exactly
uniform, the largest magnitude will bend the would-be linear trend
somewhere, and power law with small exponents can still describe the
data well.

The improved accuracy of the prediction of the main shock time of the
1989 Loma Prieta earthquake by consideration of log-periodic
oscillations was regarded as a evidence supporting the
hypothesis in \citep{sornette199505}. However, from the study presented
here, it is not rare to obtain $t_f$ near that value from our
synthetic data containing events in the range of [1940,1988]. If we
use a power law to describe the data, by definition $t_f$ should be
slightly bigger than the time of the last data point. Indeed, from our
simulations, we found that $t_f$ is distributed in [1988,1995], and
the chance of observing a synthetic $t_f$ similar to the real $t_f$ is
around 95\% of the probability of observing the most probable
synthetic $t_f$. The point is, given a sequence of events in that time
range and a power law assumption, that kind of $t_f$ is not hard to
find. 

It is important to emphasize that the present study does not alter the
usefulness of studying seismic precursors. However, the physical
interpretation associated with the observations in
\citep{sornette199505} does not seem to be warranted, at least on the
face of the Loma Prieta case only.  However, as pointed out in
\citep{huang99:_mechan_log_period_under_sampl_data} and also found in
this paper, log-periodic oscillations are robust features of power
laws. The present analysis as well as those given in
\citep{huang99:_mechan_log_period_under_sampl_data} suggests that,
whatever their origin (noise or physical), they might still be used to
improve the prediction of the main event. This is clearly what we
observe in our synthetic tests performed on pure power laws without
log-periodicity: a power law fit {\bf with} log-periodicity has a
better estimate for $t_f$ than a pure power law {\bf without} the
log-periodic oscillations.  The reason may be that, by fitting the
most probable form of noise, the fit is more stable. It seems
worthwhile to investigate this possibility further in future studies.

\appendix

\acknowledgments 
We are grateful to A. Johansen and C. Sammis for their help in retrieving
the data and to Rick Schoenberg and Y.Y. Kagan as referees for constructive
remarks.

\balance % USE THIS TO BALANCE THE LAST TWO COLUMNS IN TWOCOLUMN MODE

\bibliographystyle{agu}
\bibliography{yueqiangRef}

\end{article} % AGUTeX VERSION 4.0 AND LATER
\end{document}